\documentclass[12pt]{article}
\usepackage{amssymb, euscript}
\usepackage{amsmath,bbm}

\usepackage{theorem}
\theoremheaderfont{\scshape} 
\theoremstyle{plain}

\newcommand{\haken}{\mathbin{\hbox to 8pt{%
                 \vrule height0.4pt width7pt depth0pt
                 \kern-.4pt
                 \vrule height4pt width0.4pt depth0pt\hss}}}

\newcommand{\be}[3]{\begin{equation}  \label{#1#2#3}}
\newcommand{\bea}[3]{\begin{eqnarray}  \label{#1#2#3}}
\newcommand{\ee}{\end{equation}}
\newcommand{\ba}{\begin{array}}
\newcommand{\ea}{\end{array}}
\newcommand{\eea}{\end{eqnarray}}

\newcommand{\R}{ {\cal R} }
\newcommand{\Rh}{ \hat{\cal R} }
\newcommand{\D}{ {\cal D} }
\newcommand{\J}{ {\cal J} }

\newcommand{\M}{ {\cal M} }

\newcommand{\gent}{T\oplus T^{\ast}}

\begin{document}

\baselineskip=20pt
\parskip=6pt


\thispagestyle{empty}

\begin{flushright}
\hfill{HU-EP-04/44} \\
\hfill{hep-th/0408169}

\end{flushright}

\vspace{10pt}

\begin{center}{ \LARGE{\bf
Mirror symmetry for \\[4mm]
topological sigma models \\[4mm]
with generalized K\"ahler geometry
}}

\vspace{35pt}

{\bf Stefano Chiantese, Florian Gmeiner and Claus Jeschek}

\vspace{15pt}

{\it  Humboldt Universit\"at zu Berlin,\\
Institut f\"ur Physik,\\
Newtonstrasse 15, 12489 Berlin, Germany,\\
Email: chiantes, gmeiner, jeschek@physik.hu-berlin.de.}\\[1mm]

\vspace{8pt}

\vspace{40pt}

{\bf ABSTRACT}

\end{center}

We consider topological sigma models with generalized K\"ahler 
target spaces. The mirror map is constructed explicitly for 
a special class of target spaces and the topological A and B model 
are shown to be mirror pairs in the sense that the observables,
the instantons and the anomalies are mapped to each other.
We also apply the construction to open topological models and
show that A branes are mapped to B branes. Furthermore, we demonstrate
a relation between the field strength on the brane and a two-vector
on the mirror manifold.

\noindent

\vfill

\newpage


\section{Introduction}


Since the work of Witten \cite{Witten:1991zz} two dimensional
topological field theories have been considered to study 
mirror symmetry.
The aim of this paper is to construct explicitly
the mirror map for topological models with a certain class
of target spaces having a generalized geometry in the sense 
of Hitchin \cite{Hitchin:2002}.

Two dimensional nonlinear sigma models describe maps
$\phi : \Sigma \to X$ from the Riemann surface to the target space.
Supersymmetry on the Riemann surface constraints the target space geometry.
This link between supersymmetry and geometry becomes especially interesting
for the ${\cal N} = (2, 2)$ sigma model. In \cite{Gates:1984nk} it was realized
that the most general target space geometry is a bi-Hermitian geometry
described in terms of the following data $(g, I_+, I_-, H)$.  
The Riemannian metric $g$ is Hermitian with respect to two different
complex structures $I_+$ and $I_-$ for right and left movers, and the 
$H$-flux is a closed 3-form which can be expressed locally 
as the field strength of a $B$-field.
It was found recently \cite{Gualtieri:2004} that this geometry can be
described equivalently by
a (twisted) generalized K\"ahler structure (GKS) if ($H \neq 0$) $H = 0$.
It is defined by two commuting generalized complex
structures (GCS) $\mathcal{J}_{1/2}$, which are endomorphisms of
$\gent$, the combined tangent and cotangent spaces of the target space.
In the following we will restrict ourselves to $H=0$.

The ${\cal N} = (2, 2)$ sigma model can be twisted in two different 
ways. The twist consists in changing the spin of the fermions with 
the vector/axial symmetry so that one obtains a bosonic theory.
Twisting with the vector (axial) current, one gets the topological A (B)
model.
The twist allows to define supersymmetry also on a curved Riemann surface 
with the advantage that one can use a localization principle.

To incorporate mirror symmetry along the lines of \cite{Strominger:1996it}
into the generalized geometry picture we follow
\cite{Jeschek:2004je, Ben-Bassat:2004vn}. Work in this direction has also been
done in \cite{Fidanza:2003zi}.

In the first part of this paper
we will show how one
can define a mirror map $\M$ for a specific class of target spaces and how it
acts on the GKS. In section \ref{sec_spinorlines} we verify this 
using the language of spinor lines.
We apply these results to the generalized topological sigma models
\cite{Kapustin:2003sg,Kapustin:2004gv} described above.
We confirm that mirror symmetry transforms a generalized A model with
specific target space to a generalized B model on the mirror dual target.
It is checked explicitly in section \ref{sec_topsigma}
how the data of the sigma models (observables, instantons and $U(1)$
anomalies) transform under $\M$ and how they behave in the limit of only one
complex structure.
The generalized framework can also be applied to open topological theories
\cite{Kapustin:2003sg,Zabzine:2004dp}. In the last part of this paper we check
how mirror symmetry acts on the gluing conditions which define A and B branes.
We see explicitly how A and B branes on different target spaces are related by
$\M$.
Furthermore, we see that the two-form field is mapped to a two-vector,
which is related to a noncommutative structure on the mirror manifold.
In the appendix we give details about the construction of the generalized
topological sigma models.


\section{The mirror map $\cal M$} \label{sec-mirror}


In this section we take the GKS to be defined on a specific 6-manifold
\cite{Ben-Bassat:2004vn} (see also \cite{Jeschek:2004je}). Since
our interest is only to show how certain properties of the two GCS behave
under a specific bundle isomorphism, called mirror symmetry, we do not need
more general 6-manifolds. 

Generically, the two GCS of a GKS are given in the $\gent$ basis by
\be300
{\cal J}_{1/2}= \frac{1}{2} 
\begin{pmatrix} I_+ \pm I_- 
                 &  -(\omega_+^{-1} \mp \omega_-^{-1}) \\ 
                     \omega_+ \mp \omega_-
                 &  -(I_+^T \pm I_-^T)
\end{pmatrix},
\ee
where the complex structures $I_+$ and $I_-$ are independent
sections ($\forall p \in M^6$) in the twistor space ${\cal Z}M^6$.
Note that we always assume
integrability for the two complex structures.
We can also define a generalized metric by $G=-\J_1\J_2$.

Suppose that we take a trivial fibre bundle $M^6=T^6$ with fibre
$F=T^3$ over the 
base space $B=T^3$, thus $M^6=T^3 \oplus T^3$.
Therefore we have the following splitting of the generalized tangent space: 
\be305
(T\oplus T^{\ast})\otimes{\mathbb C} = 
     (T_B \oplus T_F \oplus T_B^{\ast} \oplus T_F^{\ast})\otimes{\mathbb C}.
\ee 

This choice is for computational convenience, but one can 
consider a more general $M^6$ as a nontrivial $T^3$ torus fibration over
a general three dimensional base space without changing the essence of 
our argument \cite{Strominger:1996it}.
Furthermore, we want to consider only GCS which are adapted in the sense
of \cite{Ben-Bassat:2004vn}, i.e.
\be310
{\cal J}_{1/2}:T_F \oplus T_F^{\ast} \to T_B \oplus T_B^{\ast}.
\ee
Respecting additionally the algebraic properties of GCS we take
\be315
I_+ \pm I_-=  
\begin{pmatrix}  &  -(\tilde I_+ \pm \tilde I_-) \\ 
                     \tilde I_+ \pm \tilde I_-  &  
\end{pmatrix},
\ee
\be320
\omega_+ \mp \omega_- = 
\begin{pmatrix} 
              &  -(\tilde\omega_+ \mp \tilde\omega_-) \\ 
                 \tilde\omega_+ \mp \tilde\omega_-  & 
\end{pmatrix}.
\ee
Note that $\tilde I_+, \tilde I_-$ and 
$\tilde\omega_+,\tilde\omega_-$ are not complex
structures and K\"ahler forms, respectively. Note also that 
to satisfy the properties $I^2_{\pm} = -1$ and $\omega^T_{\pm} = - 
\omega^T_{\pm}$ one has to require $\tilde{I}^2_{\pm} = 1$ 
and $\tilde{\omega}^T_{\pm} = \tilde{\omega}_{\pm}$.
 
We are now prepared to write the specific GCS:
\be325
{\cal J}_{1/2}= \frac{1}{2} 
\begin{pmatrix}  
      & -(\tilde I_+ \pm \tilde I_-) & & 
             -(\tilde\omega_+^{-1} \mp \tilde\omega_-^{-1})  \\ 
      \tilde I_+ \pm \tilde I_- & &  
               \tilde\omega_+^{-1} \mp \tilde\omega_-^{-1} & \\
   & -(\tilde\omega_+ \mp \tilde\omega_-)& & -(\tilde I_+^T \pm \tilde I_-^T)\\
    \tilde\omega_+ \mp \tilde\omega_- & & \tilde I_+^T \pm \tilde I_-^T &      
\end{pmatrix},
\ee
where the transpose and inverse operation only indicate that the indices
are up/down, appropriately.

We have choosen the specific manifold not by accident. We know from
\cite{Strominger:1996it} that mirror symmetry can be considered as a
T-duality transformation along the $T^3$-fibre over a
3-dimensional base space.
We adopt this idea and conjecture the mirror map $\cal M$ to be a 
map which acts on the generalized tangent bundle 
$(T\oplus T^{\ast})\otimes{\mathbb C}$
as an bundle isomorphism \cite{Ben-Bassat:2004vn,Jeschek:2004je}. 
Moreover, this isomorphism should have the property
of an involution, ${\cal M}^2 = 1$.
The mirror map in the generalized tangent space induces naturally
a map for the GKS, i.e., we get mirror transformed GCS $\hat{\cal J}_{1/2}$
and hence a mirror transformed generalized metric $\hat G$.

Let us define the mirror map such that it acts as an identity on
$T_B,T_B^{\ast}$ and as a ``flip'' on $T_F \leftrightarrow T_F^{\ast}$:
\be330
{\cal M}:T_B \oplus T_F \oplus T_B^{\ast} \oplus T_F^{\ast} \to
T_B\oplus T_F^{\ast} \oplus T_B^{\ast} \oplus T_F,
\ee
explicitly
\be335
{\cal M}=
\begin{pmatrix} 1 &  &  &  \\
         &  &  & 1 \\
         &  & 1 &  \\
         & 1 &  &  
\end{pmatrix}.
\ee
We get a conjugated GCS in the following
way:
\be340
\hat{\cal J}_{1/2}={\cal M}\circ{\cal J}_{1/2}\circ {\cal M}^{-1}:
T_B\oplus T^{\ast}_F \oplus T_B^{\ast} \oplus T_F
\to T_B\oplus T^{\ast}_F \oplus T_B^{\ast} \oplus T_F.
\ee
Applying this construction explicitly we get
\be345
\hat{\cal J}_{1/2}= \frac{1}{2} 
\begin{pmatrix}  
   &  -(\tilde\omega_+^{-1} \mp \tilde\omega_-^{-1}) & & 
         -(\tilde I_+ \pm \tilde I_-)\\
   \tilde\omega_+ \mp \tilde\omega_- & & \tilde I_+^T \pm \tilde I_-^T & \\
 & -(\tilde I_+^T \pm \tilde I_-^T) & & -(\tilde\omega_+ \mp \tilde\omega_-)\\
 \tilde I_+ \pm \tilde I_- & & \tilde\omega_+^{-1} \mp \tilde\omega_-^{-1} &  
\end{pmatrix}.
\ee
To compare $\hat{\cal J}_{1/2}$ with ${\cal J}_{1/2}$ we reinterpret 
$\hat{\cal J}_{1/2}$ as a map $T_B \oplus T_F \oplus T_B^{\ast} \oplus T_F^{\ast}
\to T_B \oplus T_F \oplus T_B^{\ast} \oplus T_F^{\ast} $ instead of
(\ref{340}). We then use the fiber metric $g_F$ and its inverse and 
we write them back into $\hat{\cal J}_{1/2}$ (see also \cite{Fidanza:2003zi}).
By using the identity $\omega = g I$, we get finally
\be355
\hat{\cal J}_{1/2}= \frac{1}{2} 
\begin{pmatrix}  
      & -(\tilde I_+ \mp \tilde I_-) & & 
             -(\tilde\omega_+^{-1} \pm \tilde\omega_-^{-1})  \\ 
      \tilde I_+ \mp \tilde I_- & &  
               \tilde\omega_+^{-1} \pm \tilde\omega_-^{-1} & \\
   & -(\tilde\omega_+ \pm \tilde\omega_-)& & -(\tilde I_+^T \mp \tilde I_-^T)\\
    \tilde\omega_+ \pm \tilde\omega_- & & \tilde I_+^T \mp \tilde I_-^T &      
\end{pmatrix},
\ee
where now $\hat{\cal J}_{1/2}$ are again maps
\be360
\hat{\cal J}_{1/2}: T_B \oplus T_F \oplus T_B^{\ast} \oplus T_F^{\ast} \to
T_B \oplus T_F \oplus T_B^{\ast} \oplus T_F^{\ast}.
\ee
This is the mirror transformed complex structure. In the following we will
denote by $\M$ the mirror map, which is the combined operation of $\M$ and
the reinterpretation of maps.
We see immediatly that mirror symmetry interchanges the two GCS:
\be365
\begin{matrix}
{\cal J}_{1/2} & \longleftrightarrow & \hat{\cal J}_{1/2}={\cal J}_{2/1}\\
(I_+, I_-) & \longleftrightarrow & (\hat I_+, \hat I_-)=(I_+, -I_-)\,.
\end{matrix}
\ee
When $M^6$ is a nontrivial torus fibration, using the same remark above,
also the mirror manifold $\hat M^6$ is a nontrivial torus fibration.

Additionally, we will derive this result by means of maximal isotropics, 
associated to pure spinor lines \cite{Hitchin:2002,Gualtieri:2004}.
But to achieve this it is usefull to
remember first the splitting of $(T\oplus T^{\ast})\otimes{\mathbb C}$
into subbundles with respect to the GCS ${\cal J}_{1/2}$.

\section{Pure spinors, maximal isotropics and the mirror map}
\label{sec_spinorlines}

Let us assume we have a generic GKS on a 6-manifold $M^6$. With the two
given commuting (integrable) GCS, ${\cal J}_{1/2}$, we get a 
decomposition of $(T\oplus T^{\ast})\otimes{\mathbb C}$ into a direct
sum of four subbundles, as it is shown in \cite{Gualtieri:2004}.
We want to review this decomposition.

On the one hand this can be 
understood by firstly noting that the generalized metric $G$ gives
a decomposition into two subbundles of dimension $3_{\mathbb C}$ each.
This can be done explicitly by using the projectors 
$P_{\pm}=\frac{1}{2}(1 \pm G)$, where the associated subbundles having eigenvalues 
$\pm1$ and carrying a positive/negative
definite metric are called $C_{\pm}\otimes{\mathbb C}$,
\be400
(T\oplus T^{\ast})\otimes{\mathbb C}=(C_+ \oplus C_-)\otimes{\mathbb C} \, .
\ee
It can be shown that elements of $C_{\pm}\otimes{\mathbb C}$ can be written as ($B=0$)
\be405
C_+\otimes{\mathbb C}=\{X + gX|X\in T\otimes{\mathbb C}\} \, ,\qquad
   C_-\otimes{\mathbb C}=\{X - gX|X\in T\otimes{\mathbb C}\} \, ,
\ee
where the generalized metric $G$ is purely Riemannian,
\be410
G=
\begin{pmatrix}
 & g^{-1}\\ g & 
\end{pmatrix} \, .
\ee
On the other hand, since the GCS commute with $G$, we can also decompose 
the generalized tangent bundle with respect to the GCS ${\cal J}_{1/2}$.
This we will do by the useful formulae,
\be415
\begin{matrix}
{\cal J}_1& = & \pi|^{-1}_{C_+}\,I_+\,\pi\, P_+ 
             + \pi|^{-1}_{C_-}\,I_-\,\pi \, P_-\,, \\
{\cal J}_2& = & \pi|^{-1}_{C_+} \, I_+ \, \pi \, P_+ 
            - \pi|^{-1}_{C_-} \, I_- \, \pi \, P_- \, ,
\end{matrix}
\ee
where $\pi:C_{\pm}\to T$ is a projection.

We will denote the $i$ eigenbundle of ${\cal J}_{1/2}$, or 
equivalently the graphs of the maps $-i{\cal J}_{1/2}$,
by $L_{1/2}$, respectively
\be420
\begin{matrix}
L_1 & = & \{X +gX|X\in T_+^{1,0} \} \oplus \{X -gX|X\in T_-^{1,0} \} \, ,\\
L_2 & = & \{X +gX|X\in T_+^{1,0} \} \oplus \{X -gX|X\in T_-^{0,1} \} \, .
\end{matrix}
\ee
The generalized tangent bundle decomposes therefore in
\be425
(T\oplus T^{\ast})\otimes{\mathbb C}=L_1\oplus\overline{L_1} 
= L_2\oplus\overline{L_2}.
\ee
Since the two GCS commute we can decompose $L_{1/2}$ further by 
${\cal J}_{2/1}$. We indicate $\pm$ for the eigenvalues $\pm i$ corresponding
to the second splitting, e.g.
\be430
L_1\oplus\overline{L_1} = L_1^+ \oplus L_1^- \oplus 
                \overline{L_1^+} \oplus \overline{L_1^-},
\ee
where
\be435
\begin{matrix}
L_1^+ & = & \{X +gX|X\in T_+^{1,0} \}\, , & \qquad &
L_1^- & = & \{X -gX|X\in T_-^{1,0} \}\, ,\\
L_2^+ & = & \{X +gX|X\in T_+^{1,0} \}\, , & \qquad &
L_2^- & = & \{X -gX|X\in T_-^{0,1} \}\, .
\end{matrix}
\ee
We see that $L_2=L_1^+ \oplus \overline{L_1^-}$ and 
\be440
C_{\pm}\otimes{\mathbb C} 
      = L_{1/2}^{\pm} \oplus \overline{ L_{1/2}^{\pm} } \,.
\ee
These careful observations make clear that by changing $I_- \to -I_-$ we
do not affect the $C_+$-bundle and moreover only interchange in the 
$C_-$-bundle holomorphic with antiholomorphic objects with respect to $I_-$.
Thus, mirror symmetry interchanges the subbundles 
$L_1^- \leftrightarrow \overline{L_1^-}$.

Now we are prepared to come back to the question: How does the mirror map
look like for pure spinor lines? We will attack this question with
our specifications for the 6-manifold given in section 2. 
For the theory of spinor lines the reader
should consult \cite{Gualtieri:2004}. Firstly, we conjecture this map
and apply it to the generating pure spinor lines. 
Let us remember that pure spinor lines single out maximal 
isotropics. More precisely, the (involutive) maximal isotropics of our interest
are ${L_1^+},{L_1^-}$. These can be described by the following four pure
spinor lines $\phi_i, \, i\in\{1,\ldots, 4\}$
\be445
\begin{matrix}
0 & = & L_1^+ \, \cdot \, \phi_1  
        &= &L_1^+ \, \cdot \, \Omega^{(3,0)}_+ \, , & \quad &
0 & =  L_1^- \, \cdot \, \phi_2 &= L_1^- \, \cdot \, \Omega^{(3,0)}_- \, , \\
0 & = & L_1^+ \, \cdot \, \phi_3 
            &=& L_1^+ \, \cdot \, e^{i\, \omega_+}\, ,  & \quad & 
0 & =  L_1^- \, \cdot \, \phi_4 &=  L_1^- \, \cdot \, e^{-i\, \omega_-} \,,   
\end{matrix}
\ee
where $\Omega^{(3,0)}_{\pm}\in \Lambda^{od}$ are holomorphic top degree forms with 
respect to $I_+,I_-$ and $\omega_{\pm} \in \Lambda^{ev}$ are the K\"ahler
forms. The operation '$\cdot$' means Clifford
multiplication: $(X + \xi)\cdot\phi = X\haken\phi + \xi\wedge\phi$, where
$X\in T,\, \xi\in T^{\ast},\, \phi\in\Lambda^{\bullet}$. 

In what follows we choose an appropriate local trivialization for our forms, i.e.
local complex coordinates with respect to either
$I_+$ or $I_-$. We split them into an imaginary part
$y^i, \, i\in\{1,2,3\}$, and a real part $x^{\alpha}, \, \alpha\in\{1,2,3\}$,
which are the coordinates in the base and the fibre, respectively. Thus, e.g.
\begin{align}
e^{i\, \omega_+} & = 
1 \, + \, i \, dx^idy^i + \, dx^{12}dy^{12}\,  +\, dx^{23}dy^{23}\,  
 + \, dx^{13}dy^{13}\,  + i \,  dx^{123}dy^{123}\,,\\
\Omega^{(3,0)}_+ & =  (dx^1 + i \, dy^1)\wedge(dx^2 + i \, dy^2)\wedge
    (dx^3 + i \, dy^3)  \,.
\end{align}
We conjecture the mirror map acting on pure spinor lines by
\begin{align}
{\cal M}: \Lambda^{ev/od} &\to \Lambda^{od/ev}\\
\phi &\to (\partial_{X_3} + dx^3)\cdot(\partial_{X_2} + dx^2)\cdot
   (\partial_{X_1} + dx^1)\cdot \phi \,,
\end{align}
where $ T_F=\text{span}\{\partial_{X_{\alpha}}\}$, 
$T^{\ast}_F=\text{span}\{dx^{\alpha}\}$ and 
$\phi\in\Lambda^{\bullet}$.

Using the property that
$\partial_{X_{\alpha}}\haken dx^{\beta}=\delta_{\alpha}{}^{\beta}$, we apply
the mirror map to our pure spinor lines $\phi_i$ to get
\begin{align}
\hat\phi_1  &= & {\cal M} \cdot \, \Omega^{(3,0)}_+  
      &= & e^{i\, \omega_+} \, ,
   &\quad&
\hat\phi_2 &=&  {\cal M} \cdot \, \Omega^{(3,0)}_- &= & e^{i\, \omega_-} \, ,\\
\hat\phi_3 &= &{\cal M} \cdot  \, e^{i\, \omega_+} 
      &=&  - \, \Omega^{(3,0)}_+\, ,
    &\quad &
\hat\phi_4 &= & {\cal M} \, \cdot \, e^{-i\, \omega_-}   
         &= & - \, \overline{\Omega^{(3,0)}_-} \, .
\end{align}
Let us now focus on the maximal isotropics which are associated to these
mirror transformed pure spinor lines $\hat\phi_i, \, i\in\{1,\ldots,4\}$. 
We see immediately that $L_1^+$ is untouched by the map ${\cal M}$ but 
in the 
$C_-$-bundle it interchanged $L_1^-$ with $\overline{L_1^-}$. Thus,
we verified exactly our previous result.

\section{Generalized topological sigma models}\label{sec_topsigma}
Topological sigma models in the generalized framework can be defined as usual
by twisting the fermionic spin with the vectorial/axial $U(1)$ charge. The crucial
difference is given by the fact that we allow for two different complex
structures. For details on the definition of the models see appendix \ref{appendix}.

We want to write the BRST operators in the $ T \oplus T^* $ bundle.
Let us define the fermionic basis
\begin{equation} \label{bigpsidef}
\psi := (\psi_+ + \psi_-) \in T , \qquad 
\rho := g(\psi_+ - \psi_-) \in T^*, \qquad 
 \Psi := \left( \begin{array}{c}
\psi \\
\rho \end{array}\right)  \, ,
\end{equation}
where $\psi_\pm$ are elements of $\pi C_\pm$.
Then the BRST operators of the generalized B and A model 
take the form \cite{Kapustin:2003sg}
\begin{eqnarray}
Q_B & = & \left< \left( \begin{array}{c}
\partial_1 \phi \\
g \partial_0 \phi \end{array} \right) ,
(1 + i {\cal J}_1 ) \Psi \right > \, , \nonumber \\[5pt]
 Q_A & = & \left< \left( \begin{array}{c}
\partial_1 \phi \\
g \partial_0 \phi \end{array} \right) ,
(1 + i {\cal J}_2 ) \Psi \right > \, ,
\end{eqnarray}
where $<\,,\,>$ is the natural metric on $\gent$ \cite{Hitchin:2002,Gualtieri:2004}.

The classical $U(1)_{A/V}$ symmetry can be broken by quantum effects. This
anomaly is given in terms of the first Chern class of the $L_{1/2}$ bundle
for the $B/A$ model \cite{Kapustin:2004gv}.
The cancellation of this anomaly constrains the target space geometry via
$c_1(L_{1/2})=0$.

Now we want to show how the relevant quantities of
the generalized B model with the target space $ M^6$ are mapped 
to the ones of the generalized A model\footnote{
This choice is of course arbitrary, one could as well start with
the generalized A model on $M^6$ and map it to the B model on $\hat M^6$}
with the mirror target space
$\hat M^6$.
{}From section \ref{sec-mirror} 
we know that ${\cal M} : {\cal J}_{1} \to {\cal J}_{2}$ so that
${\cal M} : Q_B \to Q_A$. We also know that $(I_+, I_-)$ is mapped 
to $(I_+, - I_-)$ under the mirror map and equation~(\ref{435}) 
tells us that ${\cal M} : L_1 \to L_2$. Therefore, 
${\cal M} : c_1(L_1) \to c_1(L_2) $ and the anomaly cancellation of  
the generalized B model gets mapped to that of the generalized 
A model.

The next step is to show that the observables of generalized B and A model 
are mirrors of each other. We show this for the local observables
of the closed topological sector, but first let us remember how they
were constructed in \cite{Kapustin:2003sg}. Following \cite{Witten:1991zz}, 
one has to construct scalar BRST invariant field configurations. 
Writing the BRST
variations in the $ T \oplus T^* $ bundle, we get\footnote{
This $\Phi$ is an element of $\gent$ and should not be confused with
the chiral superfield defined in appendix \ref{appendix}.
}
\be678
\delta_{B/A} \Phi  =  \Psi_{1/2} :=  \frac{1}{2}
(1 + i {\cal J}_{1/2}) 
\Psi \in  \overline{L}_{1/2}\, , \qquad \qquad \Phi :=  
\begin{pmatrix} 
\phi \\
g \phi \end{pmatrix} \, .
\ee
The nilpotency properties $\delta^2_{B/A} =0$ of the BRST variations then
yield $\delta_{B/A}\Psi_{1/2} = 0$. Thus, $\Psi_{1/2}$ are the 
configurations we are looking for in the generalized B/A model. The space of
observables is then given by
\be428
({\cal O}_f)_{B/A} = f_{a_1 \cdots a_n}(\phi) \Psi^{a_1}_{1/2} \cdots
\Psi^{a_n}_{1/2} \, ,
\ee
which can be mapped to the exterior algebra bundle 
$\Lambda^k \overline{L}_{1/2}^{\, *} \simeq \Lambda^k L_{1/2} $
since $f$ is skew symmetric in the indices $a$.
Performing the BRST variation of $({\cal O}_f)_{B/A}$, one realizes that 
the map is actually an isomorphism,
\be520
\{ Q_{B/A},  ({\cal O}_f)_{B/A} \} = 
({\cal O}_{d_{\overline{L}_{1/2}}f})_{B/A} \, , 
\ee
where $d_{\overline{L}_{1/2}} = \partial^+_{\overline{L}_{1/2}} 
+ \partial^-_{\overline{L}_{1/2}}$
is the Lie algebroid derivative such that $d_{\overline{L}_{1/2}} : C^{\infty}
(\Lambda^k L_{1/2}) \to C^{\infty}(\Lambda^{k+1} L_{1/2})$ 
\cite{Gualtieri:2004}. Since $\M : \overline{L}_1 \to \overline{L}_2$,
the cohomologies of the differential complexes for the generalized A and
B models are mirror pairs.

We want to do the same for the generalized instantons \cite{Kapustin:2003sg}.
The instantons are the fixed points of the BRST transformations. 
Performing the Wick rotation $\partial_0 \phi \to i \partial_2 \phi$
on the Riemann surface, one gets the instanton equations
\be396
\delta_{B/A} \Psi  = 
(1 - i{\cal J}_{1/2})
\begin{pmatrix} 
i \partial_2 \phi \\
g \partial_1 \phi \end{pmatrix} =
0 \, .
\ee
These equations tell us that the instantons of the generalized 
B model are mapped to those of the generalized A model under the 
mirror map. 

To conclude this section we want to show that the mirror map 
between the generalized topological models gives the old (in the sense of 
Witten \cite{Witten:1991zz}) A/B model with target space $\hat M^6$ as a mirror 
of the old B/A model with target space $M^6$. It is only necessary to note the
following. The generalized B model yields the old B and A models under the 
identifications $I_+ = I_-$ and $I_+ = - I_-$ respectively. Under the same
identifications the generalized A model yields the old A and B model.
To give just one example, the constant maps $\phi : \Sigma \to M^6$ of the old
B model are mapped to the holomorphic maps $\phi : \Sigma \to \hat M^6$
 of the old A model.

\section{Topological branes and their mirrors}

In this section we want to investigate how topological branes behave
under the mirror map $\M$.
We will strongly follow the notation and conventions used in 
\cite{Zabzine:2004dp} and
references therein.

Branes in the topological A/B model (A/B branes) can be defined by a
gluing matrix $R:T\to T$, which encodes information about the mapping of
left- and right-moving fermions at the boundary $\partial\Sigma$
\cite{Albertsson:2001dv,Albertsson:2002qc}.
The gluing conditions read
\be799
  \psi_- = R\psi_+\,.
\ee
In the generalized picture this translates to \cite{Zabzine:2004dp}
\be800
\R: \, \gent \to \gent,\qquad\qquad\R\Psi=\Psi\,,
\ee
where $\Psi$ is defined in (\ref{bigpsidef}).
$\R$ respects the natural metric $<\cdot,\cdot>$ on $\gent$,
squares to one, i.e. $\R^2=1$, and anticommutes with $G$, i.e.
$G\,\R+\R\,G=0$.

In the (physical) gluing framework the operator $\R$ contains the
information
about Dirichlet and Neumann boundary conditions (bc). It
defines a smooth distribution $\D\subset T$ which has rank equal to the
dimension of the brane. In case of an
integrable distribution we even have (Frobenius) a maximal integral
submanifold $\D$.

{}From a different point of view, the above properties of $\R$
serve to consider
the projection operator $\frac{1}{2}(1\, + \, \R)$ to
define a special almost Dirac structure $\tau^0_{\D}$
(a real, maximal isotropic sub-bundle),
\be805
\tau^0_{\D}=T\D\oplus \text{Ann}(T\D)\subset\gent \, ,
\ee
which is (Courant) integrable iff $\D$ is integrable.

The extension of $\R$ by a closed two-form
$F\in\Omega^2(\D)$, $dF=0$,  on the submanifold $\D$ corresponds to
\cite{Zabzine:2004dp}
\be807
\tau^F_{\D}=\{ \, \frac{1}{2}(1\, + \, \R)(X\, + \, \xi)=(X\, + \, \xi) \,
   : \, (X\, + \, \xi)\in T\D\oplus T^{\ast}M|_{\D} \, , \,\,
    \xi|_{\D}= X\haken F \, \}
\ee
and is equivalent to the definition of a generalized tangent bundle given
in \cite{Gualtieri:2004}.
This gluing matrix is given by
\be810
\R=\begin{pmatrix}
1 & \\ F & 1
\end{pmatrix}
\begin{pmatrix}
r &  \\ & -r^t
\end{pmatrix}
\begin{pmatrix}
1 & \\ -F & 1
\end{pmatrix}
=
\begin{pmatrix}
r & \\ F\, r\,+\, r^t\, F & -r^t
\end{pmatrix} \, ,
\ee
where $r$ is an operator which carries the gluing information for the
fermions (see also \cite{Albertsson:2001dv,Albertsson:2002qc}).

Let us focus on the A/B branes in the corresponding A/B model. This
means that the $U(1)$ currents 
$j_{\pm}=\omega_{\pm}(\psi_{\pm}$, $\psi_{\pm}), \psi_{\pm}\in T$,
have to fulfill the matching conditions
\be815
0 = j_+ \, \pm \, j_- = \frac{1}{2}\left< \Psi \, , \J_{2/1}\Psi \right >
\ee
for the A/B model, respectively.

Moreover, combining this with the gluing conditions for the fermions,
we obtain
a stability condition for $\R$, or equivalently, a stability condition for 
$\tau^F_{\D}$. Using also $\{G,\R\}=0$, one gets:
\be820
\begin{matrix}
\text{A branes:}& \R\J_1 &=& -\,\J_1\R 
              &\quad\text{and}\quad & \R\J_2 &=& \J_2\R\\
\text{B branes:}& \R\J_1 &=& \J_1\R 
               &\quad\text{and}\quad & \R\J_2 &=& -\J_2\R \, .
\end{matrix}
\ee
We will call the (anti)commuting constraints $\mp$-stability
with respect to a certain GCS.
Thus, the A/B model is $\J^{-}_{1/2}$ stable and
additionally
$\J^{+}_{2/1}$ stable. This reflects the fact that the generalized tangent
bundle $\tau^F_{\D}$ in the A/B model splits into $\pm i$ eigenbundles of
$\J_{2/1}$ or, in other words, it becomes a stable subbundle of
$L_{2/1}\oplus\overline{L_{2/1}}$, respectively:
\be822
\text{A/B model:}\quad\tau^F_{\D} \, =\,  \tau^{F\, +}_{\D}
         \oplus \tau^{F\, -}_{\D}
   \, , \quad \text{w.r.t.} \quad L_{2/1}\oplus\overline{L_{2/1}} \, .
\ee
We are now prepared to apply the mirror map $\M$ for the structures we
just
introduced. Therefore, the gluing operator $\R$ gets mapped to
$\Rh=\M \R \M^{-1}$ and one can show that the properties for $\Rh$ are
the same as they were for $\R$. As before, we take $M^6$ which has a 
$T^3$ fibration, then mirror symmetry interchanges 
Neumann bc with Dirichlet bc in the fibre\footnote{The mirror
map $\M$ is only a special case of the more general T-duality
transformation
and therefore this statement can be extended.}.
Remember that under mirror symmetry the GCS (also the A/B model)
get interchanged. Therefore
it is easy to see that the $U(1)$ current conditions get mapped to each
other
and A/B branes get naturally mapped to B/A branes. But note that on the
mirror side the stability conditions are formulated with $\Rh$.

Furthermore,
in case of non-vanishing $F\in\Omega^2(\D)$, let us
focus on the part of (\ref{810}), where only $F$ appears. Then we obtain the
following
symbolical shape of $\Rh$ 
\be825
\R=\begin{pmatrix}
r&\\ \Box&-r^t
\end{pmatrix}
\qquad\longrightarrow\qquad
\Rh=\begin{pmatrix}
\hat{r}& \Box \\ & -\hat{r}^t
\end{pmatrix}
\ee
with a bi-vector $\beta = F^{-1}$ in the upper triangular part.
Thinking of $F$ in components this means that 
the indices get raised. We will denote this by
\be835
\tau^{\beta}_{\hat\D}=\{ \, \frac{1}{2}(1\, + \, \hat\R)(X\, + \, \xi)
         =(X\, + \, \xi) \,
   :\, (X\, + \, \xi)\in T\hat M|_{\hat\D}\oplus N^{\ast}\hat\D  \, , \,
    X|_{N\hat\D}= \beta(\xi) \, \},
\ee
where $N^{\ast}\hat\D$ is the conormal bundle to $\hat\D$.
For example, if we start
with a brane which has only a worldvolume in the fibre directions and a 
non-vanishing two-form $F$, it will be mapped to a brane which corresponds 
to a ``point''. But on the mirror side $F$
disappears and we find a bi-vector $\beta$ in the fibre directions instead.
This can be interpreted as a noncommutative deformation of $\hat{M}$, as
has been argued in \cite{Kapustin:2003sg}. 


This brings us immediately to the proposal to investigate the case of
having at the same time
both independent structures, a two-form and a bi-vector.
This would correspond to $B$ and $\beta$ transformations in the sense of
\cite{Gualtieri:2004} and therefore we get a natural extension of the
generalized tangent bundle.
Until now we worked in the classical regime only. It would be very
interesting to shed more light 
(maybe along the lines of \cite{Kapustin:2003sg})
on the $\alpha'$ corrections that appear in the full string theory. A
connection to \cite{Seiberg:1999vs} should appear in this context.

\section*{Acknowledgements}

We would like to thank H. Dorn and C. Sieg for interesting discussions,
F. Witt for mathematical support and I. Runkel for proofreading the paper.

The work of the authors is supported by a Graduiertenkolleg grant of the DFG
(The Standard Model of Particle Physics - structure, precision tests and
extensions).

\appendix

\section{Appendix}\label{appendix}

\renewcommand{\t}{\theta}
\newcommand{\half}{\frac{1}{2}}
\newcommand{\ds}{\mathrm{d}^2\sigma\,}
\newcommand{\dt}{\mathrm{d}^2\theta\,}
\newcommand{\gd}[1]{\delta^{(#1)}}
\newcommand{\p}{\partial}
\newcommand{\pt}[1]{\frac{\partial}{\partial\theta^{#1}}}
\newcommand{\N}{\mathcal{N}}
\newcommand{\PP}{\mathcal{P}}
\newcommand{\wbar}[1]{\overline{#1}}

In this appendix we formulate the topological sigma models in the generalized
formalism.
We start with the nonlinear sigma model formulated in $\N=(1,1)$ superfield form:
\begin{equation}
  \label{smaction} S = \half\int\ds\dt(g+B)(D_+\Phi,D_-\Phi), \quad\mbox{where}
\end{equation}
\begin{equation}
  D_\pm = \pt{\pm}+i\t^\pm\p_\pm\quad;\quad\p_\pm:=\p_0\pm\p_1\,.
\end{equation}
The $\N=(1,1)$ SUSY transformations are generated by $Q_\pm$, defined as
\begin{equation}
  Q^{(1)}_\pm := \pt{\pm}-i\t^\pm\p_\pm\,,
\end{equation}
and the chiral superfield can be expanded in components as
\begin{equation}\label{sf} \Phi = \phi + \t^+\psi_++\t^-\psi_-+\t^-\t^+F. \end{equation}

An additional supersymmetry can be defined by \cite{Gates:1984nk}
\begin{equation} Q^{(2)}_\pm := I_\pm D_\pm. \end{equation}
This is a well defined (1,1) supersymmetry, if $I_\pm$ is a pair of integrable
almost complex structures on $T$ and $g$ is Hermitian with respect to both
$I_+$ and $I_-$. Furthermore the almost complex structures have to be
covariantly constant w.r.t. covariant derivatives with connection:
\begin{equation} \Gamma^{a}_{\pm bc} := \Gamma^{a}_{\,bc} \pm g^{ad}H_{dbc},\end{equation}
where $\Gamma$ is the Levi-Civita connection. 
We get the following relation between the two connections
\begin{equation} \label{gammarel} \Gamma^{a}_{+\,bc}\psi^b_+\psi^c_-=-\Gamma^{a}_{-\,bc}\psi^b_-\psi^c_+ \end{equation}

The variations of the superfield (\ref{sf}) can be written in components as
\begin{center}\begin{tabular}{rcl@{\hspace{2cm}}rcl}
  $\gd{1}_+\phi   $&=&$ \psi_+       $&$ \gd{1}_-\phi   $&=&$ \psi_-$\\
  $\gd{1}_+\psi_+ $&=&$ -i\p_+\phi   $&$ \gd{1}_-\psi_+ $&=&$ F$\\
  $\gd{1}_+\psi_- $&=&$ -F           $&$ \gd{1}_-\psi_- $&=&$ -i\p_-\phi$\\[3ex]
  $\gd{2}_+\phi   $&=&$ I_+\psi_+    $&$ \gd{2}_-\phi   $&=&$ I_-\psi_-$\\
  $\gd{2}_+\psi_+ $&=&$ iI_+\p_+\phi $&$ \gd{2}_-\psi_+ $&=&$ I_-F$\\
  $\gd{2}_+\psi_- $&=&$ -I_+F        $&$ \gd{2}_-\psi_- $&=&$ iI_-\p_-\phi$\\
\end{tabular}\end{center}

We can integrate out the auxiliary field $F$ using the equations of motion
\begin{equation} F^a=\Gamma^a_{+bc}\psi_+^b\psi_-^c. \end{equation}
Furthermore, we define combinations of the supersymmetry generators
\begin{equation}
\begin{array}{rclcrcl}
Q_+ & = & \frac{1}{2} (Q^{(1)}_+ - iQ^{(2)}_+) \, , & \qquad &
\wbar{Q}_+ & = & \frac{1}{2} (Q^{(1)}_+ + iQ^{(2)}_+) \, , \nonumber \\[5pt]
Q_- & = & \frac{1}{2} (Q^{(1)}_- - iQ^{(2)}_-),  &\qquad &
\wbar{Q}_- & = & \frac{1}{2} (Q^{(1)}_- + iQ^{(2)}_-) \, .
\end{array}
\end{equation}
With these we make contact to the definitions of \cite{thebook}.

We are now ready to define the generalized topological A(B) model.
We twist the spin of the fermionic fields with the vector
(axial) $U(1)$ current. The charges of the fields are given 
in the following table,
\begin{center}\begin{tabular}{l|rrrrr}
                     & $q_V$ & $q_A$ & $J$      & $J_A$ & $J_B$ \\\hline
  $\PP_+\psi_+$        & $-1$  & $-1$  & $-\half$ & $-1$  & $-1$  \\
  $\wbar{\PP}_+\psi_+$ & $+1$  & $+1$  & $-\half$ & $0$   & $0$   \\
  $\PP_-\psi_-$        & $-1$  & $+1$  & $+\half$ & $0$   & $+1$  \\
  $\wbar{\PP}_-\psi_-$ & $+1$  & $-1$  & $+\half$ & $+1$  & $0$   \\ 
\end{tabular}\end{center} 
where $q_{V/A}$ indicate the vector/axial charge.
$J$ and $J_{A/B} = J + q_{V/A}/2$ define the spins before
and after the twist and we used projectors on the (anti-)holomorphic 
parts of the fields with respect to $I_\pm$
\begin{equation} \PP_\pm=\half(1-iI_\pm),\quad \wbar{\PP}_\pm=\half(1+iI_\pm). \end{equation}

As BRST operators for the generalized A and B model we take
\footnote{Note that \cite{Witten:1991zz} uses a different definition for the
world sheet fermions, which leads to a different BRST operator for the A model,
$Q_A =Q_++\wbar{Q}_-$.}
\begin{equation} Q_A:=\wbar{Q}_+ + Q_-,\quad Q_B := \wbar{Q}_+ + \wbar{Q}_-, \end{equation}
which act on the scalar fields of the twisted models like
\begin{center}\begin{tabular}{rcl@{\hspace{2cm}}rcl}
  $\delta_A\phi$             &=& $\wbar{\PP}_+ \psi_+ + \PP_- \psi_-$ &
  $\delta_B\phi$             &=& $\wbar{\PP}_+\psi_++\wbar{\PP}_-\psi_-$ \\
  $\delta_A\wbar{\PP}_+\psi_+$ &=& $\Gamma_+\wbar{\PP}_+\psi_+\PP_-\psi_-$ &
  $\delta_B\wbar{\PP}_+\psi_+$ &=& $\Gamma_+\wbar{\PP}_+\psi_+\wbar{\PP}_-\psi_-$ \\
  $\delta_A\PP_-\psi_-$        &=& $\Gamma_-\PP_-\psi_-\wbar{\PP}_+\psi_+$ &
  $\delta_B\wbar{\PP}_-\psi_-$ &=& $\Gamma_-\wbar{\PP}_-\psi_-\wbar{\PP}_+\psi_+$ \\
\end{tabular}\end{center}

We can reformulate these variations in terms of generalized fields, written
in the $\gent$ basis, where the terms involving the connections vanish
\begin{equation} \delta_A\half(1+i\mathcal{J}_1)\Psi = 0,\quad \delta_B\half(1+i\mathcal{J}_2)\Psi = 0\,, \end{equation}
where the definition of $\Psi$ can be found in (\ref{bigpsidef}).



\providecommand{\href}[2]{#2}\begingroup\raggedright\endgroup

\end{document}